\documentstyle[12pt]{article}
\newcommand{\be}{\begin{equation}}
\newcommand{\ee}{\end{equation}}
\newcommand{\ba}{\begin{eqnarray}}
\newcommand{\ea}{\end{eqnarray}}
\newcommand{\baa}{\begin{eqnarray*}}
\newcommand{\eaa}{\end{eqnarray*}}
\newcommand{\bb}{\begin{thebibliography}}
\newcommand{\eb}{\end{thebibliography}}
\newcommand{\ci}[1]{\cite{#1}}
\newcommand{\bi}[1]{\bibitem{#1}}
\newcommand{\lab}[1]{\label{#1}}
\newcommand{\re}[1]{(\ref{#1})}
\def\stackreb#1#2{\ \mathrel{\mathop{#1}\limits_{#2}}}

\newcounter{my}
\newcommand{\he}%
   {\stepcounter{equation}\setcounter{my}%
   {\value{equation}}\setcounter{equation}0%
   }%
\newcommand{\she}%
   {\setcounter{equation}{\value{my}}%
    }%

\begin{document}
\begin{center}
{\bf SELF-SIMILARITY IN SPECTRAL PROBLEMS \\[1mm]
     AND $q$-SPECIAL FUNCTIONS
\footnote{Invited talk by the second author at the Third
International Workshop on Symmetries and Integrability of Difference
Equations (16-22 May, 1998, Sabaudia, Italy).}
}

\vspace{2mm}

{\large Igor Loutsenko$^*$ and Vyacheslav Spiridonov$^+$}

\vspace{1mm}
{$^*$ \it
     Department of Physics, Joseph Henry Laboratories, Jadwin Hall, \\
     Princeton University, P.O. Box 708,
     Princeton, NJ 08544-0708, USA; \\
 e-mail:  loutseni@feynman.princeton.edu} \\[1mm]

{$^+$ \it Bogoliubov Laboratory of Theoretical Physics, JINR,
Dubna, \\  Moscow region 141980, Russia;
e-mail: svp@thsun1.jinr.ru
}\\[5mm]
\end{center}

\begin{abstract}
Similarity symmetries of the factorization chains for one-dimensional
differential and finite-difference Schr\"odinger equations are
discussed. Properties of the potentials defined by self-similar
reductions of these chains are reviewed. In particular,
their algebraic structure, relations to $q$-special functions,
infinite soliton systems, supersymmetry, coherent states,
orthogonal polynomials, one-dimensional Ising chains and
random matrices are outlined.
\end{abstract}


\bigskip

{\bf 1. Introduction.}
Spectral problems of various types play an important role in
classical and quantum mechanics and in the theory of integrable
systems. Depending on the situation they may be based upon differential,
finite-difference or integral equations, etc. Roughly speaking
all such problems can be approximated by a search of eigenvalues of some
matrices \cite{Wilk}. This leads to a number of common features in
the methods of treatment of different spectral problems. We choose as
a basic object of discussion the ordinary one-dimensional Schr\"odinger
equation
\be\lab{eq1}
L\psi(x)=-\psi_{xx}(x)+u(x)\psi(x)=\lambda\psi(x).
\ee
The eigenvalue problem for the operator $L$ is completely defined by
imposing some boundary conditions upon $\psi(x)$. However, it is
convenient to consider \re{eq1} without such additional constraints
and assume that $x, \lambda \in C$. When a non-formal meaning
of the eigenvalues and eigenfunctions will be necessary, we assume that
$\psi(x)\in L^2(R)$.

There are several connections of \re{eq1} with finite-difference
equations and discrete systems --- the main subject of the present
workshop. First, the scheme of treatment of \re{eq1}
to be described below is universal,
e.g., it can be easily reformulated for the second order difference
equation. Second, finite-difference equations  appear already in
the consideration of \re{eq1} for different potentials. In the
space of all equations of this form finite-differences enter
through the symmetry transformations mapping Schr\"odinger
equations onto each other:
$$
u(x)\to \tilde u(x), \qquad \psi(x)\to \tilde\psi(x), \qquad
\lambda\to\tilde\lambda.
$$
Third, in some cases the discrete spectrum eigenfunctions
satisfy simple discrete equations. E.g., for $u(x)\propto x^2$
one gets from the condition of square integrability of
eigenfunctions $\lambda_n\propto n$, $\psi_n(x)\propto H_n(x)e^{-x^2/2}$,
where $H_n(x)$ are Hermite polynomials satisfying three-term recurrence
relation. Fourth, there is a direct relation of \re{eq1} to some
lattice models of statistical mechanics.

The paper is organized as follows. In the next several sections we
review briefly some old results on self-similar potentials,
their coherent states and the relation to $q$-special functions.
Then we outline briefly a generalization of the construction to
orthogonal polynomials. In the last two
sections we present our recent results on the appearance of 
one-dimensional Ising chains and random matrices in the context
of the Schr\"odinger equation \re{eq1}. The list of references
given in the end does not pretend to be complete, it contains
mostly the papers which influenced somehow the work of authors.

\medskip

{\bf 2. Self-similar spectra.}
Under the similarity transformations in $d$-dimensional Euclidean
space one counts usually the rotations, dilations and shifts of the
coordinates, $x_i\to q_{ij}x_j+a_i$, $i,j=1, \dots, d$, i.e. the
affince transformations.
In the $d=1$ case this is reduced to $x\to qx+a$, which will be
used below under the assumption that the parameters $q$ and $a$
are {\it fixed} (for $q\neq 1$ one may set $a=0$ without loss of
generality).

Let us consider separately the effects of translations and
scalings upon the Schr\"odinger operator $L$ with potential $u(x)$,
eigenfunctions $\{ \psi_j(x) \} $ and the spectrum $\{\lambda_j\}$
(the use of the subscript $j$ for notation of the spectrum is
symbolic). After application of some symmetry transformation we get
the operator $\tilde L=-d^2/dx^2+\tilde u(x)$ with the potential
$\tilde u(x)$, unnormalized eigenfunctions $\{\tilde\psi_j(x)\}$ and
the eigenvalues $\{\tilde\lambda_j\}$. So, the shift $x\to x+a$ brings in
$$
\tilde u(x)=u(x+a), \qquad \{\tilde\psi_j(x)\}=\{\psi_j(x+a)\},
\qquad \{\tilde \lambda_j\}=\{\lambda_j\}.
$$
The spectrum does not change under this transformation due to
the absence of distinguished points in the boundary conditions. In the case of
scaling transformations one replaces in \re{eq1} the variable $x$ by
$qx$, then multiplies the resulting equation by $q^2$ in order to bring
it to the canonical form and gets the operator $\tilde L$ with
$$
\tilde u(x)=q^2u(qx), \qquad \{\tilde\psi_j(x)\}=\{\psi(qx)\},
\qquad \{\tilde\lambda_j\}=\{q^2\lambda_j\}.
$$
Now one gets a scaled form of the spectrum $\lambda\to q^2\lambda$.
For $|q|>1$ the spectral points are stretched with respect to the
$\lambda=0$ point, and, vice versa, for $|q|<1$ we get a squeezed
spectrum. The complete affine transformation of the spectral
parameter requires joining of a shift $\lambda\to\lambda+ h$ which
is easily reached by the shift of the potential
$$
\tilde u(x)=u(x)+h, \qquad \{\tilde\psi_j(x)\}=\{\psi_j(x)\},
\qquad \{\tilde\lambda_j\}=\{\lambda_j+h\}.
$$
The combination of these three transformations gives
$$
\tilde u(x)=q^2u(qx+a)+h, \qquad \{\tilde\psi_j(x)\}=\{\psi_j(qx+a)\},
\qquad \{\tilde\lambda_j\}=\{q^2\lambda_j+h\},
$$
i.e. `trivial' similarity transformations of the spectral data.
The key constructive idea, bringing in a non-trivial content into the
above considerations, consists in a sticking to the systems with
{\it self-similar} spectra for which $\{\tilde\lambda_j\}=\{\lambda_j\}.$
Note that such a constraint is applicable to any spectral problem, not
just to the Schr\"odinger equation.

Analyze first the $q=1$ case. The condition $\{\lambda_j+h\}=\{\lambda_j\}$
shows that the discrete spectrum should appear in the form of a number of independent
bilateral arithmetic progressions with the increment equal to $h$. In the simplest
case the number of such progression is finite ($=N$) and one can parametrize
the eigenvalues as solutions of the equation $\lambda_j+h=\lambda_{j+N}$
or $\lambda_{pN+k}= hp+\lambda_k$, $k=1,\dots N, \; p\in Z$. The continuous
spectrum may appear in the form of arithmetic progressions of permitted bands.
For instance, if one requires that the addition of $h$ to potential is equivalent to
an isospectral transformation $x\to x+a$, i.e. $u(x)+h=u(x+a)$, then
$u(x)$ is a sum of a periodic potential $u_p(x+a)=u_p(x)$ and
the Airy part $\propto x$ known to lead to continuous spectrum filling
the whole line, $u(x)=u_p(x)+h x/a$.

Consider now the $\{q^2\lambda_i\}=\{\lambda_i\}$ case.
Formally, the discrete spectrum appears in the form of bilateral geometric
progressions accumulating near the $\lambda=0$ point from below or above.
Enumeration of pure point spectra in case of the finite number of
such progressions may be based upon the relation
$q^2\lambda_j=\lambda_{j+N} $ giving $\lambda_{pN+k}=\lambda_k q^{2p}$,
$k=1, \dots, N,\; p\in Z$.
The continuous spectrum may arise in the form of infinite number of
permitted bands concentrating near the zero.

For instance, the described situation may take place if $q^2u(qx)=u(x)$.
A solution of this equation is given by $u(x)=h(x)/x^2$, where $h(x)$
is a function periodic in the logarithmic scale $h(qx)=h(x)$.
For $h=const.$ the wave functions $\psi(x)$ are given in terms
of the Bessel functions. For $h<-1/4$ there is the ``fall onto the
center" phenomenon and under certain conditions discrete
spectrum consists of one bilateral geometric progression
of negative eigenvalues \cite{Case}. The bad point of this and of
the previous example is that the potentials $u(x)$ are not
bounded from below. Let us impose such a demand and assume that
$u(x)$ is bounded from below and non-singular. Evidently this
condition does not permit the infinite negative spectrum, i.e. it
is necessary to delete the lower tails of our geometric (or arithmetic)
progressions. However, the formula $q^2\lambda_j=\lambda_{j+N}$ still
works and gives $\lambda_{pN+k}=\lambda_kq^{2p}, p=0, 1, \dots$.
We have thus a partially self-similar spectrum:
$$
\{q^2 \lambda_j\}=\{\lambda_j\}/(\lambda_1, \dots, \lambda_N),
$$
i.e. the scaling $x\to qx$ leads to the deletion of $N$ lowest
eigenvalues from the spectrum.
Below we outline a procedure for building a class of potentials
with this property.

Note that for the $q^2\neq1$ case  the geometric progressions of
positive eigenvalues of the abstract operator $L$ are bounded from below
and no truncation is necessary. However,
one has to go beyond the Schr\"odinger equation for building
systems with such completely self-similar spectrum.

\medskip

{\bf 3. Infinite-soliton systems.}
The inverse scattering method and the general
theory of solitons (see, e.g., \cite{GGKM,Nov,AS}) provide us
a guide for construction of potentials with prefixed spectral
properties. The compatibility condition of two linear equations
$$
L\psi(x,t)=\lambda\psi(x,t), \qquad \psi_t(x,t)=A\psi(x,t)
$$
has the operator form $L_t=[A,L]$. When $L=-\partial^2_x+u(x,t)$
and $A=-\partial_x^3+6u\partial_x+3u_x$ one arrives at the KdV equation
$u_t+u_{xxx}-6uu_x=0$.

If the potential $u(x)\to 0$ for $x\to\infty$ sufficiently fast then
according to the inverse scattering method the potential $u(x)$ can
be uniquely reconstructed from the reflection coefficient $R(k)$,
discrete spectrum eigenvalues $\{\lambda_j\}$ and the
normalization constants $\{\theta_j^{(0)}\}$ for corresponding
eigenfunctions $\{\psi_j(x)\}.$ For $R(k)=0$ and finite number of
eigenvalues one arrives at the $N$-soliton potentials. The corresponding
solution of the KdV equation can be represented in the form
\be
u(x, t)=- 2\partial_x^2 \ln \tau_N(x, t),
\lab{pottau}\ee
where $\tau_N$ is the determinant of a $N\times N$ matrix $C$,
\be
\tau_N=\det C, \qquad C_{ij}=\delta_{ij}+ {2\sqrt{k_i k_j}
\over k_i+k_j} e^{(\theta_i+\theta_j)/2},
\lab{phase}\ee
$$
\theta_i=k_ix-k_i^3 t +\theta_i^{(0)}, \qquad i, j=1, 2, \dots, N.
$$
Parameters $k_i$ describe amplitudes of solitons related to eigenvalues of $L$
as $\lambda_i=-k_i^2/4$, $\theta_i^{(0)}/k_i$ are zero time
phases of solitons and $k_i^2$ are their velocities.

Let us extend this solution to the $pN$-solitonic one by imposing 
the constraint $k_{j+N}=qk_j, q<1$, and take the $p\to\infty$ limit. This leads
to the self-similar spectrum $k_{pN+m}=q^pk_m, m=1, \dots, N$, i.e. to the
infinte-soliton wavetrains whose amplitudes form $N$ geometric progressions.
The question of convergency of such a limit requires a special consideration.

Applying scaling transformations to derived potentials one
has to scale the time $t$ too, which gives $\tilde u(x,t)=q^2u(qx,q^3t)$.
The phases of this potential are $\theta_j(qx,q^3t)=
=k_{j+N}x-k_{j+N}^3t+\theta_j^{(0)}$.
There are two ``drawbacks" of the resulting expression. First, there
are infinitely many arbitrary parameters $\theta_j^{(0)}$, i.e. there
is a functional freedom in the definition of $u(x)$. Second, there is
a mixup between soliton amplitudes and phases --- the solitons with
the amplitudes $k_{j+N}$ have now the zero time phases $\theta_j^{(0)}$
instead of $\theta_{j+N}^{(0)}$. Both these shortcomings are removed
if we impose the constraint $\theta_{j+N}^{(0)}=\theta_j^{(0)}$, i.e.
$\theta_j(qx, q^3 t)=\theta_{j+N}(x,t)$, which
leads to truly self-similar infinite-soliton systems --- the dilation
$x\to qx, t\to q^3t$ just deletes $N$ solitons, corresponding to
the lowest eigenvalues of $L$. Such potential is contains 
a finite number of parameters $k_1, \dots, k_N, q, \theta_1^{(0)},
\dots, \theta_N^{(0)}$ and should be characterized by some ``finite"
equations. There are two equivalent approaches to the construction
of these equations.
The algebraic one, based upon the factorization method, and the
functional-analytic one, relying upon the Darboux transformations.

\medskip

{\bf 4. The factorization method.}
This method is well known in quantum mechanics \cite{IH}, it was
discussed already by  Schr\"odinger. Within this approach
one takes a chain of Schr\"odinger operators,
$L_j=-d^2/dx^2+u_j(x),$ and factorizes them as products of the
first-order differential operators up to some constants $\lambda_j$:
\begin{equation}
L_j=A_j^+A_j^- + \lambda_j,\qquad
A_j^\pm=\mp\;d/dx + f_j(x).
\label{e2}
\end{equation}
Then the neighboring $L_j$ are tied to each other through the
abstract factorization chain
$$
L_{j+1}=A_{j+1}^+A_{j+1}^-+\lambda_{j+1}=A_j^-A_j^++\lambda_j,
$$
i.e. one passes from $L_j$ to $L_{j+1}$ just by the permutation
of operator factors. This gives the intertwining relations
$A_j^-L_j=L_{j+1}A_j^-,\; L_jA^+_j=A_j^+L_{j+1}. $
As a result, if one has $L_j\psi^{(j)}=\lambda\psi^{(j)}$, then
$\psi^{(j+1)}\propto A_j^-\psi^{(j)}$. Indeed,
$L_{j+1}(A_j^-\psi^{(j)})=A_j^-L_j\psi^{(j)}=\lambda(A_j^-\psi^{(j)}).$
Under particular conditions the action of the operator $A_j^-$ leads to
removing or addition of an eigenvalue in the spectrum of the operator
$L_{j+1}$ with respect to the spectrum of $L_j$. There may take place
an isospectral situation as well. If zero modes of all $A_j^-$ are
normalizable, then $\lambda_j, \lambda_{j+1}, \dots$ form the discrete
spectrum of $L_j$.

Let us define the operators
$$
M_j^-=A_{j+N-1}^-\dots A_{j+1}^-A_j^-, \qquad
M_j^+=A_j^+A_{j+1}^+\dots A_{j+N-1}^+.
$$
They relate eigenfunctions of the operators $L_j$ and $L_{j+N}$
due to the intertwining relations
$$
L_{j+N}M_j^-=M_j^-L_j, \qquad M_j^+L_{j+N}=L_jM^+_j.
$$
It is easy to see that the products of operators $M^\pm_j$ should
commute either with $L_j$ or with $L_{j+N}$. Indeed, one has
the equalities
\be
M_j^+M_j^-=\prod_{k=0}^{N-1}(L_j-\lambda_{j+k}), \qquad
M_j^-M_j^+=\prod_{k=0}^{N-1}(L_{j+N}-\lambda_{j+k}),
\lab{polrel}\ee
which, together with the previous relations, look almost as an
algebra of symmetries.

The main advantage of this method is that it is not tied
to any particular type of spectral problem. One may realize
the abstract factorization chain with the help of differential,
difference, integral, etc operators.

\medskip

{\bf 5. Darboux transformations.} One starts from the semi-disrecte $LA$-pair
$$
L_j\psi^{(j)} = \lambda\psi^{(j)}, \qquad L= -d^2/dx^2 + u_j(x),
$$
$$
\psi^{(j+1)} = A_j^-\psi^{(j)}, \qquad A_j^- = d/dx + f_j(x).
$$
The compatibility condition yields the intertwining relation
$A_j^-L_j = L_{j+1}A_j^-$, the resolution of which gives
$u_j = f_j^2 - f_{jx} + \lambda_j,  u_{j+1} = u_j + 2f_{jx} = f_j^2 + f_{jx} +\lambda_j$.
The substitution
$f_j = -\phi_x^{(j)}/\phi^{(j)} $ converts the relation between $u_j$ and $f_j$
into the equation $-\phi_{xx}^{(j)} + u_j\phi^{(j)} = \lambda_j\phi^{(j)}. $
As a result one gets the original Darboux transformation:
$\psi^{(j+1)} = \psi^{(j)}_x - (\phi^{(j)}_x/\phi^{(j)})\psi^{(j)}.$

The compatibility condition (or the factorization chain) is equivalent to the following 
infinite chain of nonlinear differential-difference equations
\be
\left(f_j(x)+f_{j+1}(x)\right)_x+f_j^2(x)-f_{j+1}^2(x)=\mu_j\equiv\lambda_{j+1}-\lambda_j.
\lab{dchain}\ee
One may search solutions of this equation in the form of power series in $j$.
The finite term expansion occurs only if $f_j(x)=a(x)j+b(x)+c(x)/j$,
where $a, b, c$ are some elementary functions of $x$ \cite{IH}.
The resulting Schr\"odinger equation is solved in terms of the $_2F_1$
hypergeometric function.

\medskip

{\bf 6. Polynomial supersymmetry.}
Upon the two neighboring operators $L_j$, $L_{j+1}$ of the
factorization chain a simple superalgebra describing the
boson-fermion symmetry is realized. Let us introduce
the $2\times 2$ matrix Hamiltonian $H=-d^2/dx^2+f_j^2(x)-f_{jx}(x)\sigma_3$
and define the supercharges
$Q^\pm=A_j^\pm (\sigma_1\pm i\sigma_2)/2$
($\sigma_k$ are the Pauli matrices). Then one has
$\{Q^+, Q^-\} =H, \; [H, Q^\pm]=(Q^\pm)^2=0.$
This construction was generalized in \cite{RS} to a
symmetry between particles with parastatistics (para-supersymmetry).
In the simplest case one takes the Hamiltoninan given by
$3\times 3$ diagonal matrix with the entries $L_{j-1}, \; L_j,\; L_{j+1}$.
A symmetry between the boson and a parafermion of the second order
is described now by some polynomial (cubic) relations between $H$ and
the correspodning generalized supercharges.

As a next step in the ``nonlinearization" of supersymmetry the following
polynomial supersymmetry algebra has been derived in \cite{AIS}
$$
\{Q^+, Q^-\}= P(H), \qquad [H, Q^\pm]=(Q^\pm)^2=0,
$$
where $Q^\pm=M_1^\pm (\sigma_1\pm i\sigma_2)/2$ and
$P(H)$ is a polynomial of $H$:
$$
P(H)=\prod_{k=1}^{N}(H-\lambda_k), \qquad
H=\left(\begin{array}{cc}
 L_1 & 0 \\
 0   & L_{N+1}
\end{array}\right).
$$
Since the supercharges have now more than one zero mode,
the Witten index does not characterize the supersymmetry
breaking. One may deform this algebra via the affine
transformations and impose various natural constraints upon
$H$. This gives an alternative way of derivation (which we
are not describing here) of self-similar potentials
considered in the next section.

\medskip

{\bf 7. Self-similar potentials.}
The infinite soliton potentials with self-similar spectra considered above
appear as self-similar solutions of the chain (\ref{dchain}).
The simplest $N=1$ case has been considered in \cite{Sh,S}.
The general class of these potentials is defined by the following $q$-periodic
reduction \cite{S1}:
\begin{equation}
f_{j+N}(x)=qf_j(qx),\qquad \mu_{j+N}=q^2\mu_j.
\label{e8}
\end{equation}
Note that the solitonic interpretation is valid only if $0<q^2<1$
and there are no singularities on the whole line $-\infty < x <\infty,$
which is not true for arbitrary $q, x, \lambda_j\in C$.
What are the properties of these potentials ?

{\it Algebraically}, these are the systems whose symmetries
are described by the quantum algebras. Indeed, the equations
\re{polrel} form a closed algebra in the
cases when the operators $L_j$ and $L_{j+N}$ are related to
each other in some way, e.g., if they are similar to each other $L_j\sim
L_{j+N}$. The $q$-periodic closure corresponds to the
following constraint $L_{j+N}=q^2TL_jT^{-1}$, where $T$ is
the scaling operator $T\psi(x)=\sqrt{|q|}\psi(qx)$. This gives 
$u_{j+N}(x)=q^2u_j(qx)$.

For real $q\neq 0,\; T$ is the unitary operator $T^\dagger=T^{-1}$.
Substituting the operator constraint into \re{polrel} one
arrives at the following polynomial algebra
\begin{equation}
LB^\pm=q^{\pm 2}B^\pm L, \qquad
B^+B^-=\prod_{k=1}^{N}(L - \lambda_k),\qquad
B^-B^+=\prod_{k=1}^{N}(q^2L - \lambda_k),
\label{e14}
\end{equation}
where we have denoted $L\equiv L_1, B^-\equiv T^{-1}M_1,
B^+\equiv M_1^+T$.
For $N=1$ this is a $q$-analog of the harmonic oscillator algebra
\ci{Mac},  for $N=2$ one gets the $su_q(1,1)$ algebra, etc.

{\it Analytically}, we deal with a class of functions appearing
from solutions of the following system of nonlinear
differential-$q$-difference equations:
$$
(f_1(x)+f_2(x))_x+f_1^2(x)-f_2^2(x)=\mu_1, \quad \dots\dots
$$
$$
(f_N(x)+qf_1(qx))_x+f_N^2(x)-q^2f_1^2(qx)=\mu_N.
$$

A flavor of the structure of the general solution is obtained
from consideration of various limiting cases, when $f_j(x)$
are expressed through known functions. The simplest situation is
obtained when $f_j(x)=c_j=const.$ and arbitrary $q$. This leads
to $L=-d^2/dx^2$, i.e. to the free nonrelativistic particle, which
is interpreted in this way as a $q$-algebraic system. When $f_1(x)$
is not singular at $x=0$, the crystal base limit $q\to 0$ results in
the general $N$-soliton potential. One can take the limit $q\to 1$
in such a way that the geometric progressions in the spectrum are
converted into the arithmetic ones. Then for $N=1,2$ one gets the
potentials $u(x)\propto x^2, ax^2+b/x^2$. For $N=3,4$ the functions
$f_j(x)$ are expressed through the Painlev\'e IV and V transcendents
\cite{VS,Adler}. For $q=-1$ one gets a similar situation but
now the functions are constrained to obey certain structure under the
parity transformation. It is possible to consider the limit from
Schr\"odinger equation to classical mechanics, in which
case the self-similar potentials are determined by the much more simple
functions \ci{sym,EK}.

For $q$ a primitive root of unity, $q^n=1,
q\neq \pm1$, one gets the finite gap potentials with additional
(quasi)crystallographic symmetries \cite{Skorik}. The appearance of
Painlev\'e functions and of the finite-gap potentials in a similar
setting was discussed also in \ci{Weiss,Fl}. It is natural to
refer to the whole class of functions emerging for general 
initial conditions and $0<|q|<1$ as the continuous $q$-Painlev\'e
functions. Because of the emergence of quantum algebras, the derived
class of self-similar functions represents a new set of $q$-special
functions defined upon the differential equations. They differ in
structure from the basic hypergeometric series \cite{GR}, but
there is a connection with them via the coherent states.

Let us remark that the Darboux transformations allow one to
remove or add only a finite number of levels from/to the spectrum of a given
Schr\"odinger operator. If one finds an intertwining operator deleting
or adding permitted bands from/to the given spectrum, then
the systems with a self-similar infinite gap
spectrum can be constructed from the requirement that the action
of this intertwining operator is equivalent to a simple dilation of
the coordinate $x$.

\medskip

{\bf 8. Coherent states.}
There are many definitions of coherent states in physics.
The purely group-theoretical approach uses the orbits
of groups lying behind the symmetries of a taken physical system.
For simple spectrum generating algebras, as in
the harmonic oscillator or singular oscillator cases, these
states may be defined as eigenfunctions of the lowering operator.
In this approach coherent states play the role of
generating functions of irreducible representations of the
underlying algebra. Below we outline briefly the results of
application of the latter definition to the nonlinear algebra
of symmetries \re{e14}. A more detailed discussion of the
structure of these coherent states can be found in \ci{S2}.

The action of operators $B^\pm$ upon abstract eigenstates
of the operator $L$, $L|\lambda\rangle=\lambda|\lambda\rangle,$ 
has the form:
$$
B^-|\lambda\rangle=
\prod_{k=1}^N\sqrt{\lambda-\lambda_k}|\lambda q^{-2}\rangle, \qquad
B^+|\lambda\rangle=
\prod_{k=1}^N\sqrt{\lambda q^2-\lambda_k}|\lambda q^{2}\rangle.
$$
Let $0<q<1$ and $\lambda_k<0$. Then the $\lambda<0$
eigenstates of $L$ are formed from  up to $N$ 
lowest weight unitary irreducible
representations of the algebra \re{e14}.
Indeed, $B^-$ is the lowering operator for $\lambda<0$ states and
$\sqrt{\lambda-\lambda_k}$
becomes complex for $\lambda<\min \{\lambda_k\}$. Since 
$B^-|\lambda_k\rangle=0$, this problem does not arise
for a special choice of $\lambda$. Namely, if the point $\lambda<0$
belongs to the spectrum of $L$, it must be of the form
$\lambda_{pN+k}\equiv \lambda_kq^{2p}$
and the corresponding state is $|\lambda_{pN+k}\rangle \propto
\left(B^+\right)^p|\lambda_k\rangle$ (we assume that $|\lambda_k\rangle$
are normalizable).

Coherent states for these series are defined as
eigenstates of the lowering operator $B^-$:
\begin{equation}
B^-|\alpha\rangle_-^{(k)}=\alpha|\alpha\rangle_-^{(k)},\quad k=1, \dots, N.
\label{coh1}\end{equation}
There are $N$ such states since they are defined for each lowest weight
series separately. Representing $|\alpha\rangle_-^{(k)}$ as a superposition
of the states $|\lambda_{pN+k}\rangle$ one finds
$$
|\alpha\rangle_-^{(k)}\propto \sum_{p=0}^\infty C_p^{(k)}\alpha^p
|\lambda_{pN+k}\rangle\propto {_N}\varphi_{N-1}\left(
{0, \dots, 0 \atop b_1^k, \dots, b_{N-1}^k}; q^2, z\right)|\lambda_k\rangle,
$$
where ${_N}\varphi_{N-1}$ is a basic hypergeometric series
with the operator argument $z=(-1)^N\alpha B^+/\lambda_1\dots\lambda_N$
and the parameters $\{q^2, b_j^k\}=\{q^2\lambda_k/\lambda_j\}$. 
The general definition of the series of this type is \cite{GR}
$$
_r\varphi_s \Biggl( {a_1, \; a_2,\; \dots, \; a_r \atop
b_1,\; b_2,\; \dots, \; b_s}; q, z\Biggr)=
\sum_{n=0}^\infty {(a_1, a_2, \dots, a_r; q)_n\over
(q, b_1, \dots, b_s; q)_n}[(-1)^nq^{n(n-1)/2}]^{1+s-r} z^n,
$$
where $r$ and $s$ are arbitrary positive integers, and $a_1, \dots, a_r,
b_1, \dots, b_s$ are free parameters. We use also the 
following compact notations $(a;q)_n=(1-a)\dots(1-q^{n-1})$ and 
$(a_1, a_2, \dots, a_n; q)_n=(a_1; q)_n(a_2; q)_n\dots (a_n;q)_n.$
These coherent states are normalizable 
if the complex variable $\alpha$ lies inside of the circle
$|\alpha|^2< |\lambda_1\dots\lambda_N|$.
General $_r\varphi_s$ series are related to coherent states of
the rational generalization of the algebra \re{e14} \ci{Od}.

The $\lambda=0$ eigenstates of $L$ are simultaneously eigenstates of
the $B^\pm$ operators, i.e. they have a direct interpretation as
coherent states. This is a special degenerate representation
of the algebra \re{e14}.

Unusual coherent states appear from the non-highest
weight representations of (\ref{e14}) corresponding to the $\lambda>0$
eigenstates of $L$. Let $E_0>0$ be a discrete spectrum point.
Then the action of $B^\pm$ generates discrete spectrum of $L$ in the form
of one bilateral geometric progression $E_0 q^{2n}, n\in Z$
(such situation cannot take place for the Schr\"odinger equation).
In principle there may be an {\it arbitrary}
number of such progressions because $\sqrt{\lambda-\lambda_k}$
is real for arbitrary $\lambda>0$ and there are no truncation
conditions as in the $\lambda<0$ case.

Since for $\lambda>0$ the lowering operator is $B^+$,
the coherent states should be defined as eigenstates of this operator
(instead of $B^-$):
\begin{equation}
B^+|\alpha\rangle_+=\alpha|\alpha\rangle_+.
\label{coh2}\end{equation}
These coherent states can be expanded over the series of eigenstates
of $L$ with positive eigenvalues $|E_0q^{2n}\rangle$:
$$
|\alpha\rangle_+\propto \sum_{n=-\infty}^\infty C_n\alpha^n|E_0q^{2n}
\rangle \propto {_0\psi_N}\left({0, \dots, 0\atop \lambda_1/E_0, \dots,
\lambda_N/E_0}; q^2, z \right)|E_0 \rangle,
$$
where ${_0\psi_N}$ is a bilateral $q$-hypergeometric series
with the operator argument $z=\alpha B^-/(-E_0)^N$.
The general representative  of these series is defined as follows
\cite{GR}:
$$
_r\psi_s\left({a_1, \dots, a_r \atop b_1, \dots, b_s}; q, z\right)=
\sum_{n=-\infty}^\infty {(a_1, \dots , a_r; q)_n \over
(b_1, \dots, b_s; q)_n} \left((-1)^nq^{n(n-1)/2}\right)^{s-r}z^n.
$$
Coherent states $|\alpha\rangle_+$ are normalizable if
the parameter $\alpha$ lies outside of the previous indicated
circle $|\alpha|^2>|\lambda_1\dots\lambda_N|$.

When the $\lambda>0$ region is occupied by continuous spectrum, coherent states
are defined as eigenfunctions of $B^+$ again, but in this case it is necessary to use
integrals in the expansion of $|\alpha\rangle_+$ over
the states $|\lambda\rangle$. Since the continuous spectrum may be considered
as a continuous direct sum of irreducible representations with bilateral
geometric progressions formed by spectral points, there are now infinitely many
coherent states. Let $N=1$ and the states $|\lambda\rangle$ are not degenerate.
Then, normalizing $\lambda_1=1/(q^2-1)$, we have
$$
|\alpha\rangle_+^{(s)}= C(\alpha)
\int_0^\infty \frac{\lambda^{\gamma_s}|\lambda\rangle d\lambda}
{\sqrt{(-\lambda q^2(1-q^2);q^2)_\infty}},
$$
where
$$
\gamma_s=\frac{2\pi i s-\ln(\alpha q^2\sqrt{1-q^2})}
{\ln q^2}, \qquad   s\in {\bf Z}.
$$
These states are normalizable for $|\alpha|^2 > 1/(1-q^2),$ and have the
unit norm for
$$
|C|^{-2}=\int_0^\infty\frac{\lambda^{-\tau}d\lambda}{(-\lambda q^2(1-q^2);q^2)_\infty}
=\frac{\pi}{\sin\pi\tau}\frac{(q^{2\tau};q^2)_\infty (q^2(1-q^2))^{\tau-1}}{(q^2;q^2)_\infty},
$$
where $\tau=\ln|\alpha q\sqrt{1-q^2}|/\ln q.$
The last integral is calculated exactly being a particular subcase of
a Ramanujan $q$-beta integral \cite{GR}.

Schr\"odinger operators with self-similar potentials have the continuous
spectrum for $\lambda>0$. It 
is doubly degenerate, which leads to duplication of
the number of coherent states. An instructive example
is provided by the free nonrelativistic particle model for
which $L=-d^2/dx^2$. The corresponding $q$-harmonic oscillator
algebra generators have the form 
$$
B^-=T^{-1}(d/dx+1/\sqrt{1-q^2}),\qquad B^+=(-d/dx+1/\sqrt{1-q^2})T,
$$
$$
B^-B^+-q^2B^+B^-=1, \qquad L=B^+B^- - 1/(1-q^2).
$$
The eigenvalue
problem $B^-\psi_\alpha^-(x)=\alpha\psi_\alpha^-(x)$ leads to the
pantograph equation, which was analyzed in detail in \ci{KaM}.
The corresponding results show that in accordance
with the purely algebraic consideration there are no normalizable
coherent states of this type. However, there are
infinitely many normalizable eigenstates of the operator $B^+$
determined as appropriate solutions of the advanced pantograph
equation \ci{KaM}
$$
\frac{d}{dx}\psi_\alpha^+(x)=-\alpha q^{-3/2}\psi_\alpha^+(q^{-1}x)
+\frac{q^{-1}}{\sqrt{1-q^2}}\psi_\alpha^+(x).
$$
An important property of these states is that they are determined
by $C^\infty$ but not analytical functions at the $x=0$ point.
Consideration of the $N>1$ symmetry algebras in this realization
leads to the generalized pantograph equations \ci{Is}.

As a general conclusion to this section we would like to stress that
all three types of objects associated with $q$-special functions
in their standard meaning \ci{GR}: the ordinary and bilateral basic
hypergeometric series, and Ramanujan type integrals, show up in the
context of self-similar spectral problems.

\medskip

{\bf 9. Second order difference equation.}
Let us describe briefly a realization of the algebra of symmetries \re{e14}
upon the discrete Schr\"odinger equation or  three term
recurrence relation for orthogonal polynomials \cite{SVZ,SZ2,S3}.
One considers an infinite chain of Jacobi matrices $L_j$ and the
corresponding eigenvalue problems:
\be
L_j\psi_n^j\equiv \psi_{n+1}^j+u_n^j\psi_{n-1}^j+b_n^j\psi_n^j
=\lambda\psi_n^j, \qquad n, j\in Z.
\label{dse}\ee
In the case of orthogonal polynomials, one considers \re{dse} only for $n>0$ and 
imposes the boundary conditions $\psi_0^j(\lambda)=0,$ $\psi_1^j(\lambda)=\lambda-b_1^j$.
Simlar to the continuous case the forward discrete time step
is defined by the Christoffel's transformation to kernel polynomials \ci{Sz}:
\be
\psi_n^{j+1}=\frac{\psi_{n+1}^j+C_n^{j+1}\psi_n^j}{\lambda-\lambda_{j+1}}
\equiv \frac{S_{j+1}\psi_n^j}{\lambda-\lambda_{j+1}}.
\label{forwdse}\ee
The backward transformation was analyzed by Geronimus \ci{Ger} and it has the form:
\be
\psi_n^{j-1}=\psi_n^j+A_n^j\psi_{n-1}^j\equiv R_j\psi_n^j.
\label{backdse}\ee
Here $A_n^j$ and $C_n^j$ are discrete analogs of the superpotentials $f_j(x)$
and $S_j, R_j$ are the first order difference operators.
The compatibility conditions of the $j\to j\pm 1$ moves yield the factorizations
$L_j=S_jR_j+\lambda_j$, $L_{j-1}=R_jS_j+\lambda_j$, or 
$u_n^j=A_n^jC_n^j, \; b_n^j=A_{n+1}^j+C_n^j+\lambda_j.$
The abstract factorization chain in this case is equivavlent
to the following set of nonlinear finite-difference equations:
\be
A_n^{j}C_{n-1}^{j}=A_n^{j-1}C_n^{j-1}, \quad
A_n^{j}+C_n^{j}+\lambda_{j} = A_{n+1}^{j-1} +C_n^{j-1}+\lambda_{j-1},
\label{factdse}\ee
known to define a discrete-time Toda lattice. Note that this is not an isospectral
flow since the constants $\lambda_j$ determine the character of the change of the
spectrum of the taken Jacobi matrix after the Christoffel or Geronimus transformations.

Discrete Schr\"odinger equation analogs of the self-similar reductions \re{e8} were
described in \cite{SVZ}:
\be
A_n^{j+N}=qA_{n+k}^j, \qquad C_n^{j+N}= qC_{n+k}^j, \qquad
\lambda_{j+N}=q\lambda_j,
\label{qperdse}\ee
where $k$ is an integer. When $n$ is considered as continuous, $k$ may be taken as a
continuous variable as well.
This closure is associated with classical, semi-classical and, so-called, Laguerre-Hahn
class of orthogonal polynomials on linear and $q$-linear grids \cite{Mag}.
The corresponding recurrence
coefficients are related to ordinary and $q$-analogs of some discrete Painlev\'e transcendents
\ci{SVZ,NPR}. The algebra of symmetries of these systems is derived along the same lines as 
in the continuous case, however, in the present case the compact versions of algebras,
like $su_q(2)$, are allowed as well. The coherent states are defined in the same way as
in the continuous case. For some explicit examples see, e.g., \ci{ASu}.

The following discrete-time Volterra lattice has been derived in \ci{SZ2}:
\be
D_n^j \left( D_{n-1}^j-\beta_j \right)=D_n^{j-1}\left( D_{n+1}^{j-1}
- \beta_{j-1} \right).
\label{dtvl}\ee
It can be mapped upon the discrete-time Toda lattice \re{factdse} via the
following quadratic relation:
\be
A_n^j=D_{2n}^jD_{2n+1}^j, \qquad
C_n^j=(D_{2n+1}^j-\beta_j)(D_{2n+2}^j-\beta_j),
\label{map1}\ee
$$
\lambda_j=const. -\beta_j^2,
$$
which generalizes well-known relation between the ordinary Toda and Volterra lattices.
There is also the second similar mapping
\be
A_n^j=D_{2n-1}^jD_{2n}^j, \qquad
C_n^j=(D_{2n}^j-\beta_j)(D_{2n+1}^j-\beta_j),
\lab{map2}\ee
with the same connection between $\lambda_j$ and $\beta_j$ as given before.
As shown in \cite{SZ2}, this discrete-time Volterra lattice is related to the $g$-algorithm
proposed by Bauer in numerical analysis \ci{Bau}.
There exists an ansatz of semi-separation
of discrete variables in \re{dtvl} which leads \ci{SZ2} to recurrence coefficients of the
Askey-Wilson polynomials --- the most general set of classical orthogonal polynomials
\cite{AW} and, simultaneously, of the polynomials considered by Askey
and Ismail \cite{AI}.

An interesting discrete symmetry for the chain \re{dchain} has been described
in \ci{Adler}. It is associated with a freedom in the intermediate steps of
two-step discrete-time shifts $j\to j+2$ and related to the statement on
permutability of a sequence of B\"acklund-Darboux transformations for fixed
set of corresponding parameters. In \ci{S3} an analog of this symmetry for the
discrete time Toda and Volterra lattices has been derived.

Let us describe briefly this refactorization symmetry for the equation
(\ref{dtvl}). Let $\beta_j\neq 0$ for any $j$, then:
\be
\tilde D_{n}^j=\frac{1}{\beta_{j-1}}\left( \beta_jD_{n}^j +
{(\beta_j^2-\beta_{j-1}^2)(D_n^jD_{n+1}^j+D_n^{j-1}D_{n+1}^{j-1}) \over
\beta_j(\beta_j-D_{n-1}^j-D_{n+1}^j) +
\beta_{j-1}(\beta_{j-1} - D_n^{j-1}-D_{n+2}^{j-1})}
\right),
\lab{dnj}\ee
\be
\tilde D_{n}^{j-1}=\frac{1}{\beta_j} \left( \beta_{j-1}D_n^{j-1} -
{(\beta_j^2-\beta_{j-1}^2)(D_{n-1}^jD_n^j+D_{n-1}^{j-1}D_n^{j-1})
\over \beta_j(\beta_j-D_{n-2}^j-D_n^j)+
\beta_{j-1}(\beta_{j-1}-D_{n-1}^{j-1}-D_{n+1}^{j-1})
}\right).
\lab{dnj-1}\ee
The change of spectral parameters $\beta_j$ looks as follows:
\be
\tilde \beta_j=\beta_{j-1}, \qquad \tilde\beta_{j-1}=\beta_j.
\lab{betalaw}\ee
If one substitutes into \re{dtvl} instead of $D_n^j, D_n^{j-1}$ and 
$\beta_j, \beta_{j-1}$ the tilded variables and keeps
all other $D_n^k, \beta_k, k\neq j, j-1,$ fixed, then the
resulting equation will be satisfied automatically, i.e. we have a 
discrete symmetry.
The transformation laws for superpotentials $A_n^j, C_n^j$, describing a
similar symmetry for the
discrete-time Toda lattice, follow from the maps \re{map1}, \re{map2}.
If $\beta_j=0$ or $\beta_{j-1}=0$ ($j$ is fixed), then there appears some additional
freedom, we refer for details to \cite{S3}.

We have sketched only the simplest types of discrete-time integrable
systems. For an analysis of ordinary Toda lattice and its self-similar
reductions see, e.g., \ci{Levi,LW}. For more complicated examples and
their applications see \ci{Hir0,NRK,KLWZ,ND}. Recently self-similar 
reductions of spectral transformation chains have been considered for 
biorthogonal rational functions in \ci{SZ4}.

\medskip

{\bf 10. One-dimensional Ising chains.}
Tau-function of the $N$-soliton solution of the KdV equation \re{phase} can be represented
in the following Hirota form, which was widely
discussed in the literature (see, e.g., \cite{Sol,AS}
and references therein):
\begin{equation}
\tau_N = \sum_{\mu_i=0,1} \exp \left( \sum_{1\leq i<j \le N } A_{ij}
\mu_i \mu_j + \sum_{i=1}^N \theta_i\mu_i \right). \label{N_soliton}
\end{equation}
Here the phase shifts $A_{ij}$ are expressed via the spectral variables $k_i$ in 
a simple way
\begin{equation}
e^{A_{ij}}={ (k_i-k_j)^2 \over  (k_i+k_j)^2 }.  \label{KDV_phase}
\end{equation}
In \ci{LS} it was noticed that this expression for $\tau_N$ defines the grand
partition function of the lattice gas model if
$\theta_i=\theta^{(0)}=const$. A comparison with \cite{Bax}
shows that $\mu_i$ play the role of the filling factors of the lattice sites
by molecules and $\theta^{(0)}$ is a chemical
potential. The constants $A_{ij}$ are proportional to the interaction energies
of molecules.

Simultaneously, $\tau_N$ describes the partition function of a
particular one-dimensional Ising chain:
\be
Z_N=\sum_{\sigma_i=\pm1} e^{- \beta E}, \qquad \beta=\frac{1}{kT},
\lab{Zising}\ee
$$
E =\sum_{1\leq i<j \leq N} J_{ij}\sigma_i\sigma_j -\sum_{i=1}^N H_i\sigma_i,
$$
where $N$ is the number of spins $\sigma_i=\pm 1$, $J_{ij}$ are  the
exchange constants, $H_i$ is an external magnetic
field, $T$ is the temperature and $k$ is the Boltzmann constant.
Indeed, after the substitution into \re{N_soliton} of the spin variables
$\mu_i=(\sigma_i+1)/2$, some simple calculations yield
\be
\tau_N=e^{\Phi} Z_N, \qquad \Phi=\frac{1}{4}\sum_{i<j} A_{ij}
+ \frac{1}{2}\sum_{j=1}^N \theta_j,
\lab{tauising}\ee
with the following relations between soliton parameters and Ising chain characteristics
\be
\beta J_{ij} =-\frac{1}{4} \; A_{ij},
\qquad
\beta H_i=\frac{1}{2}\theta_i +\frac{1}{4}\sum_{j=1, i\neq j}^N A_{ij}.
\lab{identif}\ee

A similar relation with Ising chains is valid for the whole 
KP hierarchy and some other
differential and difference nonlinear integrable equations.
The corresponding tau-functions have the form \re{N_soliton} for different
choice of the phase
shifts $A_{ij}$ and the phases $\theta_i$, a list of such equations
can be found, e.g., in \ci{Sol,AS}.

A crucial observation of \ci{LS} is that the condition of self-similarity
of the spectrum is related to translational invariance of the spin
exchange constants $J_{ij}$ of the infinite Ising chain induced by the KdV
equation. In the simplest case  one demands translational invariance
of the system with respect to the shift by one site $j\to j+1$ which means
that $J_{i+1,j+1}=J_{ij}$. As a result, the exchange $J_{ij}$ or phase shifts
$A_{ij}$ depend only on the distance between the sites $|i-j|$.
This natural physical requirement forces $k_i$ to form
one geometric progression
\begin{equation}
k_i=k_1q^{i-1}, \qquad
q=e^{-2\alpha}, \qquad A_{ij}=2\ln|\tanh\alpha (i-j)|,
\label{interactions} \end{equation}
where $k_1$ and $q<1$ are free parameters.
One may demand also the invariance with respect to shifts by a multiple of
the lattice site, when $J_{i+M, j+M}=J_{ij}$, and this leads to the general
self-similar spectra $k_{j+M}=qk_j$.

In fact the translational invariance in not exact for
finite chain $1\leq j \leq N$. The limit $N\to\infty$
corresponds to the thermodynamic limit. It gives an infinite
soliton potential with self-similar discrete spectra.
The coordinate $x$ and time $t$ (and higher ``times"  of the corresponding
hierarchies) are interpreted as parameters of the magnetic field $H_i$.
The $x, t$ dependent part of $H_i$ decays exponentially
fast for $i\to\infty$, since $q<1$.
Therefore in the thermodynamic limit only the values of constants
$\theta_i^{(0)}$ are relevant for the partition function.
The formalism allows us to treat the $M$-periodic magnetic fields,
$H_{i+M}=H_i$, which for $M=1$ is just a homogeinity condition.

For the KdV equation case one has $0<|\tanh \alpha(i-j)|<1$
and $J_{ij} = -A_{ij}/4\beta > 0$, i.e. an antiferromagnetic Ising chain.
A similar situation takes place for general $M$-periodic case.
Such interaction has a long distance character but the intensity falls
off exponentially fast. The absence of phase transitions in such systems
for nonzero temperatures is well known.

For $\alpha\to 0$ or $q\to 1$
the phse shifts $A_{ij}\propto J_{ij}/kT$ are diverging.
If one renormalizes the exchange constants
$J_{ij}^{ren}=J_{ij}(q^{-1}-q)$ and the temperature
$kT_{ren}=kT(q^{-1}-q)$, then the interaction energy of a
single spin with all other ones will be finite for $q\to 1$.
As a result, in this limit one actually gets a nonlocal interaction
model with a low effective temperature.
Note that for imitation of the change
of the temperature it is necessary to change simultaneously the
magnetic field $H=H^{ren}/(q^{-1}-q)$.

The limit $q \to 0$ gives $J_{ij}^{ren}\propto \delta_{i+1,j}$,
or the high temperature nearest neighbor interaction Ising chain
(if $H$ is renormalized). If $H$ is kept finite then this limit
corresponds to the non-interacting spins. So, the formalism
provides only a partial description of the partition
function in a two-dimensional subspace of $(T, H, q)$.
For fixed $q$ the temperature $T$ has a prescribed value and
one may normalize the ``KdV temperature'' to $\beta=1$.

Using the Wronskian formula for the representation of $N$-soliton
potential \ci{Crum}, we were able to calculate the partition function
for the translationally invariant Ising chains in a homogeneous magnetic field.
Omitting technical details, which can be found in \ci{LS}, for $M=1$ 
we get $Z_{N}\to \exp(-N\beta f_I)$ for $N\to\infty$,
where the free energy per site $f_I$ has the form
\be
-\beta f_I(q, H)=\ln \frac{2(q^4;q^4)_\infty \cosh \beta H}{(q^2;q^2)_\infty^{1/2}}
+ \frac{1}{4\pi}\int_0^{2\pi}d\nu \ln (|\rho(\nu)|^2 - q\tanh^2 \beta H),
\lab{free}\ee
$$
|\rho(\nu)|^2=
\frac{(q^2e^{i\nu};q^4)_\infty^2(q^2e^{-i\nu};q^4)_\infty^2}
{(q^4e^{i\nu};q^4)_\infty^2(q^4e^{-i\nu};q^4)_\infty^2}\;
\frac{1}{4\sin^2(\nu/2)}=
q\frac{\theta_4^2(\nu/2, q^2)}{\theta_1^2(\nu/2, q^2)}.
$$
Here $\theta_{1,4}(y, q)$ are the standard Jacobi theta-functions of
the argument $y$ and base $q$ (our base is $q^2$).
The original expression for the density function $\rho(\nu)$ is
\begin{equation}
\rho(\nu)=\frac{(q^2;q^4)_\infty^2}{(q^4;q^4)_\infty^2}
\sum_{k=-\infty}^\infty
\frac{e^{i\nu k-\epsilon k}}{1-q^{4k+2}}=
\frac{(q^2e^{i\nu-\epsilon};q^4)_\infty (q^2e^{-i\nu+\epsilon};q^4)_\infty}
{(e^{i\nu-\epsilon};q^4)_\infty (q^4e^{-i\nu+\epsilon};q^4)_\infty},
\label{nu}
\end{equation}
where a small regularization parameter $\epsilon\to 0$
is introduced for absolute convergency of the infinite sum.
The last equality in \re{nu} was obtained with the help of the
$_1\psi_1$ Ramanujan's bilateral
basic hypergeometric series sum \ci{GR}:
$$
\sum_{n=-\infty}^\infty \frac{(a;q)_n}{(b;q)_n} z^n
= \frac{(q,b/a,az,q/az;q)_\infty}{(b,q/a,z,b/az;q)_\infty}.
$$
This fact demonstrates again the general statement given earlier
that the $q$-special functions are always behind the self-similar
spectral problems.

The derivative of $f_I(H)$ with respect to $H$ yields
the total magnetization of the lattice:
$$
m(H)=-\partial_H f_I = \stackreb{\lim}{N\to\infty} \frac{1}{N}
\sum_{i=1}^{N}\langle \sigma_i \rangle
$$
\be
=\tanh \beta H\left(1-\frac{1}{\pi}
\int_0^{\pi}\frac{\theta_1^2(\nu, q^2)d\nu}{\theta_4^2(\nu, q^2)\cosh^2\beta H
-\theta_1^2(\nu, q^2) \sinh^2\beta H} \right).
\lab{magkdv}\ee
The function $m(H)$  grows monotonically with $H$, its
qualitative properties are predicted by the general
theorems on  the behavior of 1D systems with the fast decaying
interactions.

The main drawback of the described scheme is that
the exact calculation is valid only for a fixed temperature in the
partition function ($q$ is fixed). One may
try to replace (\ref{interactions}) by
$A_{ij}=2 n\ln|(k_i-k_j)/(k_i+k_j)|,$ where $n$ is a sequence of
integers, to change the temperature and look for integrable models
with such phase shifts. This is not a simple task since there
are many constraints involved into the resolution of this problem.
We have found only one more example \ci{LS} for $n=2$
corresponding to a subcase of the $N$-soliton solution of the
KP-equation of B-type (BKP). The
BKP equation has the form (see, e.g., \cite{DJKM,Hir2})
\be
{\partial \over \partial x_1} \left( 9{\partial u \over \partial x_5}
-5 {\partial^3 u \over \partial x_3 \partial x_1^2}+{\partial^5 u \over
\partial x_1^5} - 30{\partial u \over \partial x_3}{\partial u \over
\partial x_1}+
30{\partial u \over \partial x_1}{\partial^3 u \over \partial x_1^3}+
60\left({\partial u \over \partial x_1}\right)^3 \right)
 - 5{\partial^2 u \over \partial x_3^2}=0.
\label{BKPeq}\ee
$\tau$-function of this integrable equation
generates partition functions of quite general Ising chains.
The exchange constants have now the form
\begin{equation}
\beta J_{ij}=-\frac{1}{4}A_{ij}, \quad e^{A_{ij}}=\frac{(a_i-a_j)(b_i-b_j)(a_i-b_j)(b_i-a_j)}
{(a_i+a_j)(b_i+b_j)(a_i+b_j)(b_i+a_j)}
\label{BKP}\end{equation}
with the energy and partition function given by (\ref{Zising}).
For $a_i=b_i=k_i/2$ one gets the KdV-inspired model for $\beta=n=2$.

Translational invariance of the spin lattice,
$J_{ij}=J(i-j)$,
results in the following spectral self-similarity
\begin{equation}
a_i=q^{i-1}, \quad b_i=bq^{i-1}, \quad q=e^{-2\alpha},
\label{BKP_momentums}
\end{equation}
where we set $a_1=1$ and assume that $\alpha >0$.
This gives the exchange
$$
\beta J_{ij}=-\frac{1}{4}\ln\frac{\tanh^2\alpha(i-j)- (b-1)^2/(b+1)^2}
{\coth^2\alpha(i-j) - (b-1)^2/(b+1)^2}.
$$
Since this expression is invariant with respect to the inversion
$b\to b^{-1}$, the parameter $b$ is restricted to the
unit disk $|b|\leq 1$. It is not difficult to see that for
$-1 < b < -q$ one has now the ferromagnet, i.e. $J_{ij}<0$.
For two other physical regions $q< b \leq 1$ and $b=e^{i\phi}\neq -1,$
one has the antiferromagnet, i.e. $J_{ij}>0$.

In the thermodynamic limit $N \to \infty $ the partition function
can be represented as the determinant of a Toeplitz matrix,
which is diagonalized by the discrete Fourier transformation.
Using again the $_1\psi_1$ sum, we have found 
the free energy per site for the homogeneous magnetic field $H_i=H$:
$$
-\beta f_I(H)=\frac{1}{4}\ln\frac{(q, q, bq, q/b; q)_\infty}
{(-q, -q, -bq, -q/b; q)_\infty} +\frac{1}{4\pi}\int_0^{2\pi}d\nu \ln|2\rho(\nu)|,
$$
where
\be
\rho(\nu)=\cosh 2\beta H
+\frac{(-q;q)_\infty^2}{(-e^{i\nu}, -qe^{-i\nu}; q)_\infty}
\left(
\frac{(b^{-1}e^{i\nu}, qbe^{-i\nu}; q)_\infty}{(b^{-1}, qb; q)_\infty}
+\frac{(be^{i\nu}, qb^{-1}e^{-i\nu}; q)_\infty}{(b, qb^{-1}; q)_\infty}
\right).
\lab{free_bkp}\ee
The magnetization if obtained after taking the derivative with respect to $H$
\be
m(H) = \tanh 2\beta H\;
\left(1-\frac{1}{\pi}\int_0^\pi \frac{d\nu}
{1+d(\nu)\cosh 2\beta H}\right),
\lab{magbkpgen}\ee
where
$$
d(\nu)=\frac{(qb,q/b;q)_\infty (b^{-1/2}-b^{1/2})\theta_2(\nu, q^{1/2})}
{(-q;q)_\infty^2 2\mbox{Im }\theta_1(\nu-(i/2)\ln b, q^{1/2})}
$$
for $q<b\leq 1$,
$$
d(\nu)=\frac{(qb,q/b;q)_\infty (|b|^{-1/2}+|b|^{1/2})\theta_2(\nu, q^{1/2})}
{(-q;q)_\infty^2 2\mbox{Re }\theta_2(\nu-(i/2)\ln|b|, q^{1/2})}
$$
for $-1<b<-q$, and
$$
d(\nu)=\frac{(qe^{i\phi},qe^{-i\phi};q)_\infty 2\sin(\phi/2)\;\theta_2(\nu, q^{1/2})}
{(-q;q)_\infty^2 [\theta_1(\nu+\phi/2, q^{1/2})-\theta_1(\nu-\phi/2, q^{1/2})]}
$$
for $b=e^{i\phi}$. Here $\theta_2(\nu, q^{1/2})$ is another Jacobi
$\theta$-function. Taking the limit $b\to 1$ one may find the magnetization
for the second value of discrete temperature of the KdV equation inspired
spin chain, $n=2$. The details of calculations and graphical representation 
of $m(H)$ for some values of the parameters can be found in \ci{LS}.  
% As expected the magnetizations for ferromagnetic cases are higher for 
% the antiferromagnetic ones. 
In a somewhat similar way one may consider the $M>1$ periodic systems as well.

\medskip

{\bf 11. Random matrices.}
Random matrices were employed by Wigner and Dyson
for studying spectra of complex systems with
many degrees of freedom. When the density of
levels is high enough, excitation energies
can be described statistically.
The detailed structure of hamiltonians is not known and
statistical characteristics of the
system are described by averaging over ensembles of random matrices.
A fundamental constraint is that the probability distributions should
be invariant under basic symmetries such as parity, rotation and
time-reversal transformations \cite{Meh}.

Gaussian ensembles use a real symmetric, Hermitian or
self-dual Hermitian random matrix $H$
with the statistically independent elements $H_{ik}$.
The probability density $P(H)$ for $H_{ik}$ to lie in a unit volume
is proportional to
$\exp\left(-a \, {\rm tr} \, H^2 + b \, {\rm tr} \, H\right),$
with the measure invariant under orthogonal, unitary or symplectic
transformations. Abandoning the condition of statistical independence
one may pass to ensembles with more complicated $P(H)$.

Dyson has introduced circular ensembles of unitary random $n\times n$
matrices $S$ with eigenvalues $\epsilon_j=e^{i\phi_j}$, $j=1, \dots, n$,
such that the behavior of the phases $\phi_i$ is equivalent to the
distribution of eigenvalues of a system. The condition of
invariance of the measure under all unitary authomorphisms
$S \to USW,$ where $U, W$ are arbitrary unitary matrices,
determines the probability distribution.
Any $S$ can be represented in the form $S=U^{-1}EU,$
$E$ a diagonal matrix with the eigenvalues $\epsilon_j$ and
$ U $ some unitary matrix.  Then, using the invariance of probability measure,
one can gauge away the $U$-dependent part of $S$ and integrate out the
``angular variables" in the probability distribution.
As a result, the eigenvalue distribution becomes
$$
Pd\phi_1 \dots d\phi_n\propto \prod_{i<j}|\epsilon_i-\epsilon_j|^2 d\phi_1 \dots d\phi_n.
$$
One can weaken the condition of invariance of the measure under
arbitrary left or right translations and leave only pure unitary
transformations. Then the invariant measure is not uniquely defined and 
the eigenvalue distribution may be written as
$$
Pd\phi_1 \dots d\phi_n=f(\phi_1,\dots,\phi_n)\prod_{i<j}|\epsilon_i-\epsilon_j|^2 d\phi_1 \dots d\phi_n,
$$
where the function $f$ is symmetric under permutation of its arguments.
One of such ensembles has been considered by Gaudin long time ago \cite{Gaud1}.
In this model one has the following probability law for eigenvalues
\begin{equation}
Pd\phi_1 \dots d\phi_n\propto
\prod_{i<j}\left|\frac{\epsilon_i-\epsilon_j}
{\epsilon_i-z\epsilon_j}\right|^2 d\phi_1 \dots d\phi_n,
\label{distribution1}
\end{equation}
interpolating between the Dyson unitary ensemble ($z=0$) and the uniform
distribution ($z=1$).

The same model can be interpreted also as a Coulomb gas on a circle
with the partition function
$$
Z_n\propto \int_0^{2\pi}\dots\int_0^{2\pi}d\phi_1\dots d\phi_n
\exp\left(-\beta \sum_{i<j}V(\phi_i-\phi_j)\right),
$$
where $\beta=1/kT $ is fixed and the potential energy is
\begin{equation}
\beta
V(\phi_i-\phi_j)=\ln\left(1+\frac{\sinh^2\gamma}{\sin^2((\phi_i-\phi_j)/2)}\right),
\quad z=e^{-2\gamma}.
\label{gas1}
\end{equation}

The grand partition function of this model corresponds to the
$\tau$-function of the KP hierarchy in the specific infinite
soliton limit \ci{LS}. The finite soliton solutions provide thus
a discretization of the system, namely, a lattice gas on the
circle.
% with equal spacing
%between the sites. Such a model is equivalent to the Ising spin chain on the circle in homogeneous
%magnetic field, the exchange constants being given by (\ref{gas1}).
For example, one may take unitary $n\times n$ matrices $S$ with the
eigenvalues equal to $N$-th roots of unity, i.e.
$\phi_j=2\pi m_j/N, m_j=1, \dots, N $.
The measure is taken to be continuous in the ``angular" variables and
discrete in eigenvalues, i.e. one takes the sums over $\phi_j$ instead
of the integrals. The completely continuous model
is recovered for $N\to\infty$, $\phi_j$ fixed:
\begin{equation}
\left(\frac{2\pi}{N}\right)^n\sum_{m_1=1}^N \dots \sum_{m_n=1}^N
\stackreb{\to}{N\to\infty} \int_{0}^{2\pi}d\phi_1 \dots \int_{0}^{2\pi}d\phi_n.
\label{sum}\end{equation}
For $n\times n$ matrices the lattice partition function is
$$
Z_n(N,z)=\left(\frac{2\pi}{N}\right)^n\sum_{m_1=1}^N \dots
\sum_{m_n=1}^N \prod_{1 \le i < j \le n} \left|\frac{\epsilon_i-\epsilon_j}
{\epsilon_i/\sqrt{z}-\sqrt{z}\epsilon_j}\right|^2,
$$
$$
\epsilon_j=\exp{\frac{2i\pi m_j}{N}}, \qquad z=e^{-2\gamma}.
$$
The grand canonical ensemble partition function is
\begin{equation}
Z(z,\theta)=\sum_{n=0}^N \frac{Z_n(N,z)e^{\theta n}}{n!} =
\sum_{\mu_m=0,1} \exp \left( \sum_{1 \le m<k \le N}A_{mk}\mu_m\mu_k+
(\theta+\eta) \sum_{m=1}^N \mu_m \right),
\label{grand}\end{equation}
where $ \eta=\ln({2\pi}/{N}) $ enters as an addition to the chemical
potential $ \theta $ and % (cf. with \re{gas1})
$$
A_{mk} = \ln \frac{\sin^2(\pi(m-k)/N)}{\sin^2(\pi(m-k)/N) +\sinh^2\gamma}
= \ln \frac{(a_m-a_k)(b_m-b_k)}{(a_m+b_k)(b_m+a_k)},
$$
is the KP phase shift with the restricted choice of parameters
\begin{equation}
a_m=e^{2i\pi m/N}, \qquad b_m=-z a_m, \quad m=1, 2, \dots, N.
\label{momentums}\end{equation}
So, the grand partition function of this discrete circular unitary matrix model
coincides with the particular KP  $N$-soliton $\tau$-function at zero 
hierarchy ``times".
In the thermodynamical limit $ N \to \infty $ the relation (\ref{sum}) takes place
and we get the matrix model \re{distribution1}. In December 1998 we have
known that this root-of-unity discretization was considered by
Gaudin himself \ci{Gaud2}, where the connection with
Ising chains was noticed as well but the relation with integrable
equations was not established.

The BKP equation suggests a further generalization of this matrix model
with the probability law
\begin{equation}
Pd\phi_1 \dots d\phi_n\propto
\prod_{i<j}\left|\frac{\epsilon_i-\epsilon_j}
{\epsilon_i+\epsilon_j}\right|^2\left|\frac{\epsilon_i+z\epsilon_j}
{\epsilon_i-z\epsilon_j}\right|^2 d\phi_1 \dots d\phi_n.
\label{distribution2}\end{equation}
The grand partition function is also defined by (\ref{grand}) where
$A_{mk}$ is given by \re{BKP}. 
In order to escape singularities appearing from zeros
$\epsilon_i+\epsilon_j=0$ for $\phi_j=\phi_i +\pi$,
the parameters $a_m, b_m$ have to be
restricted. The choice \re{momentums} works for odd
$N$ only, which makes the $N\to\infty$ limit problematic.
Another option is to replace
$m/N$ in \re{momentums} by $m/2N$, which corresponds
to the half-circle system. Comparing the connection of BKP
equation with Ising chains described in the previous section
we see, that now we have spins on the circle,
$q$ is a root of unity and the exchange $J_{mk}=J(m-k)$ is
$$
\beta J(m) =-\frac{1}{4}\ln\frac{1+\sinh^2\gamma/\cos^2(\pi m /N)}
{1+\sinh^2\gamma/\sin^2(\pi m/N)}.
$$

An important difference between the $|q|<1$ and $|q|=1$ Ising
chains is that in the former case the thermodynamic limit
$N\to\infty$ does not lead necessarily to the continuous space
models as in the circular case. Also the relation of non-compact
Ising chains to hermitian random matrices is not so straightforward.
In particular, in order to get nontrivial weight functions
one has to take inhomogeneous magnetic field corresponding to
nonzero hierarchy ``times". Calculation of the partition function
is more involved in this situation.

\medskip

{\bf 12. Conclusions.}
Self-similarity in spectral problems serves as a key to special functions,
in particular, to $q$-special functions. We hope that the brief review
given in this paper gives a sufficient number of arguments for convincing
in this vague statement.
As to the notion of exact solvability of a given equation, it requires
a more rigorous group-theoretical definition in terms of the operator
(differential, difference, etc) Galois theory. In this approach
one fixes in advance the basic fields of functions
entering spectral problems and asks about the transcendence of solutions
of the resulting equations over these fields. In the scheme we have presented the
situation is slightly different --- the basic fields of functions are defined as
solutions of complicated nonlinear equations appearing after self-similar
reductions of some infinite-dimensional systems of equations. Usually this gives
simultaneously a part of solutions of the underlying linear spectral equations
in quadratures over the basic field of functions, but
determination of the trancendence of the general solution remains open. Therefore
these two approaches should be used in conjunction. One of the problems appearing
on this route is to give a complete characterization of the changes in the
structure of corresponding (differential, \dots) Galois groups resulting from the 
Darboux-type spectral transformations \ci{S2}.

Self-similar spectra considered here are based upon
linear map of the spectral parameter $\lambda\to q\lambda+h$.
This leads naturally to spectra containing a finite number of
geometric progressions and ``Painlev\'e-type" transcendental
functions on the background.
A natural generalization involves polynomial
maps $\lambda\to P(\lambda)$ leading in general to chaotic structure
of the spectrum. The situation now is much more complicated and
only the simplest possibilities were considered in the literature.
Note that, similar to the linear case, superpositions of several
spectral sequences generated by the map $\lambda\to P(\lambda)$
are permitted too, but strong difficulties arise in the
characterization of functions involed in these systems already
in the simplest cases (see, e.g., \ci{BGH,BM}). The scalings of discrete
grids, or decimations entering at this stage hint upon
the relevance of the self-similarity hidden in wavelets and
quasicrystals. Perhaps all these subjects are naturally unified
within an extended appoach to the concept of self-similarity.

\medskip

{\it Acknowledgments.}
The authors are indebted to many people for interest to
the results reviewed in this paper, stimulating discussions 
and a collaboration, in particular, to
R. Askey, Yu. Berest, R. Hirota, A. Iserles, K. Kajiwara,
A. Orlov, T. Shiota, S. Skorik, S. Suslov, C. Tracy, L. Vinet and A. Zhedanov.
V.S. is greatful to D. Levi and O. Ragnisco
for inviting to the SIDE III workshop and a kind hospitality.
The first author is supported by a fellowship from NSERC (Canada).
The second
author is partially supported by the RFBR (Russia) grant 97-01-00281 and
the INTAS grant 96-0700.

\end{document}